\input phyzzx.tex
\tolerance=1000
\voffset=-0.0cm
\hoffset=0.7cm
\sequentialequations
\def\rl{\rightline}

\def\t1{{\tilde 1}}

\def\t{\theta}

\REF{\DYN}{E. Poppitz and S. Trivedi, Ann. Rev. Nucl. Part. Sci. {\bf 48} (1998) 307, [arXiv:hep-th/9803107]; Y. Shadmi and Y. Shirman, Rv. Mod. Phys. {\bf 72} (2000) 25, 
[arXiv:hep-th/907225]; K. Intriligator and N. Seiberg, Class. Quant. Grav. {\bf 24} (2007) S741, [arXiv:hep-th/0702069].}
\REF{\MU}{I. Antoniadis, E. Gava, K. S. Narain and T. R. Taylor, Nucl. Phys. {\bf B432} (1994) 187, [arXiv:hep-th/9405024]; A. Brignole, L. Ibanez and C. Munoz, Phys. Lett. {\bf B387} (1996)
769, [arXiv:hep-ph/9607405]; P. Nath and T. R. Taylor, Phys. Lett. {\bf B548} (2002) 77, [arXiv:hep-ph/0209282].}
\REF{\BCW}{R. Blumenhagen, M. Cvetic and T. Weigand, Nucl. Phys. {\bf B771} (2007) 113, [arXiv:hep-th/0609191].}
\REF{\IA}{L. Ibanez and A. Uranga, JHEP {\bf 0703} (2007) 052,[arXiv:hep-th/0609213]; JHEP {\bf 0802} (2008) 103, arXiv:0711.1316[hep-th].}
\REF{\IR}{L. Ibanez and R. Richter, JHEP {\bf 0903} (2009) 090, arXiv:0811.1583[hep-th].}
\REF{\CHR}{M. Cvetic, J. Halverson and R. Richter, arXiv:0905.3379[hep-th].}
\REF{\INS}{R. Blumenhagen, M. Cvetic and S. Kachru, T. Weigand, arXiv:0902.3251[hep-th] and references therein.}
\REF{\GW}{D. Green and T. Weigand, arXiv:0906.0595[hep-th].} 
\REF{\GEO}{C. Vafa, Journ. Math. Phys. {\bf 42} (2001) 2798, [arXiv:hep-th/0008142]; F. Cachazo, K. Intriligator and C. Vafa, Nucl. Phys. {\bf B603} (2001) 3, [arXiv:hep-th/0103067].}
\REF{\DIN}{M. Dine, J. Feng and E. Silverstein, Phys. Rev {\bf D74} (2006) 095012, [arXiv:hep-th/0608159].}
\REF{\SHA}{M. Aganagic, C. Beem and S. Kachru, Nucl. Phys. {\bf B796} (2008) 1, arXiv:0709.4277[hep-th].}
\REF{\SUP}{E. Witten, Nucl. Phys. {\bf B507} (1997) 658, [arXiv:hep-th/9706109]; M. Aganagic and C. Vafa, [arXiv:hep-th/0012041].}
\REF{\QUI}{F. Cachazo, S. Katz and C. Vafa, [arXiv:hep-th/0108120].}
\REF{\NPS}{S. Gukov, C. Vafa and E. Witten, Nucl. Phys. {\bf B584} (2000) 69, Erratum-ibid, {\bf B608} (2001) 477, [arXiv:hep-th/9906070].}

\singlespace
\rl{SU-ITP-09/26}
\pagenumber=0
\normalspace
\medskip
\bigskip
\titlestyle{\bf{An Exponentially Small $\mu$--term in String Theory}}
\smallskip
\author{ Edi Halyo{\footnote*{e--mail address: halyo@stanford.edu}}}
\smallskip
\centerline {Department of Physics} 
\centerline{Stanford University} 
\centerline {Stanford, CA 94305}
\smallskip
\vskip 2 cm
\titlestyle{\bf ABSTRACT}

We describe a mechanism that produces an exponentially small $\mu$--term on the world--volumes of D5 branes wrapping a deformed and fibered $A_3$ singularity. The small $\mu$ arises due
to brane instanton effects which can be calculated after a geometric transition at one of the nodes of the singularity.

\singlespace
\vskip 0.5cm
\endpage
\normalspace

\centerline{\bf 1. Introduction}
\medskip

The hierarchy problem, namely the smallness of the Higgs mass compared to the Planck (or some other high) scale and its stability under quantum corrections, is one of 
the celebrated problems in physics. The Higgs mass has to be around the $TeV$ scale in order to obtain acceptable electro--weak symmetry breaking without fine tuning.
It is well--known that supersymmetry can stabilize the Higgs mass around the supersymmetry breaking scale (taken to be $\sim TeV$) solving the stability problem. 
This leaves the $TeV$ scale tree level Higgs mass to be explained.
In the Minimally Supersymmetric Standard Model (MSSM), Higgs masses 
get two separate contributions; one from the soft supersymmetry breaking terms and the other from the supersymmetry preserving $\mu$ parameter, i.e. the term $\mu H_u H_d$ in the superpotential.
(Here $H_u,H_d$ are the Higgs doublets that couple to the up and down quarks respectively.) $TeV$ scale soft supersymmetry breaking scalar masses can be naturally derived from
dynamical supersymmetry breaking[\DYN]. Nonperturbative effects which drive dynamical supersymmetry breaking are exponentially small (compared to the Planck scale) and lead to small 
soft scalar masses. However, since the $\mu$--term
preserves supersymmetry, supersymmetry breaking cannot explain its small size. In fact, $\mu$ is the only dimensional parameter that appears in the MSSM superpotential and 
there is no reason why it should be suppressed relative to the Planck scale. Such a large $\mu$ would be a problem since it can lead to acceptable electro--weak symmetry breaking
only with extreme fine tuning (assuming equally large soft scalar masses). This is the $\mu$--problem and is another manifestation of the hierarchy problem.

A possible solution to the $\mu$--problem is to postulate a term in the superpotential like
$$W=\lambda \phi H_u H_d \eqno(1)$$
where $\phi$ is a new (singlet) field beyond the MSSM spectrum and $\lambda$ is a coupling of $O(1)$.
Then, if $\phi$ gets a $TeV$ scale VEV, the $\mu$--problem would be solved. Of course, as it stands, this is not a solution to the problem since it is just as hard to obtain
a $TeV$ scale VEV for $\phi$ as it is to obtain such a small $\mu$. On the other hand, one may couple $\phi$ to a hidden sector non--Abelian gauge group 
and hope to get an exponentially small VEV from its nonperturbative dynamics. Even though such a toy model is relatively easy to build in field theory it seems quite arbitrary. Thus, it is 
important to find out if string theory provides a better solution to the $\mu$--problem along these lines.

In this letter, we propose a possible solution to the $\mu$--problem in string theory[\MU-\CHR]. We first describe a simple mechanism in field theory that involves a new field (as in eq. (1))  
and solves the $\mu$--problem. We then realize a similar mechanism in string theory. We consider D5 branes wrapped on a deformed $A_3$ singularity fibered over $C(x)$. The world--volume 
theory of these D5 branes
includes doublets that we identify with the Higgs doublets and other singlets that play the role of the singlet $\phi$. The superpotential for these fields include Yukawa, mass and F terms.
A nonperturbative correction to the superpotential arises due to brane instantons[\INS], namely from
Euclidean D1 branes wrapping one of the the $S^2$s that defines the deformation of the singularity. This effect can be calculated classically after a geometric transition[\GEO] that changes
the geometry by $S^2 \to S^3$. The low energy superpotential of the world--volume theory, after the heavy fields are integrated out, results in a $\mu$--term that is exponentially small 
due to the D1 instanton effects.{\footnote1{While this paper was prepared but could not be submitted to the arXiv [\GW] appeared which is along similar lines.}}
We show that without fine tuning and for reasonable values of parameters, a $TeV$ scale $\mu$ can be obtained.

The letter is organized as follows. In the next section, we describe a simple mechanism in field theory that gives an exponentially small $\mu$. In section 3, we describe the brane model
that does the same and discuss why it is a better solution than the one in field theory. Section 4 includes a discussion of our results and our conclusions.

\bigskip
\centerline{\bf 2. A Mechanism for Small $\mu$ in Field Theory}
\medskip

In this section we describe a simple model in field theory that solves the $\mu$--problem by retrofitting[\DIN]. We assume that
there is a singlet, $\phi$, that couples to both Higgs bosons as in eq. (1). We would like to build an effective theory that gives an exponentially small VEV to $\phi$. Consider the
nonrenormalizable superpotential
$$W=m \phi^2+ Tr(W_{\alpha} W_{\alpha}) \left(a{\phi \over M_P}+b{\phi^2 \over M_P^2}+\ldots \right) \eqno(2)$$
where $W_{\alpha}$ are superfields for a hidden non--Abelian gauge group which condenses at a high scale, $\Lambda$, (with $TeV<< \Lambda << M_P$).
$a$ and $b$ are numerical constants of $O(1)$ and the dots in eq. (2) denote the higher order nonrenormalizable terms. 
Strong dynamics of the hidden gauge group causes gauginos to condense with $<Tr(W_{\alpha} W_{\alpha})> \sim \Lambda^3$.
Due to the superpotential in eq. (2), gaugino condensation 
leads to F and mass terms for $\phi$ where
$$F \sim a {\Lambda^3 \over M_P} \qquad m_{np} \sim b{\Lambda^3 \over M_P^2} \eqno(3)$$
Since the superpotential contains a tree level mass for $\phi$ and $\Lambda<<M_P$ we can neglect the nonperturbative contribution to the mass. However, since there is no tree level F--term,
the contribution to the F--term in eq. (3) cannot be neglected. The effective superpotential becomes
$$W=m \phi^2+ F \phi \eqno(4)$$
where $F$ is given by eq. (3). Using $V=|\partial W/ \partial \phi|^2$ we get
$$V=|2m \phi+F|^2 \eqno(5)$$
Clearly this potential has a supersymmetric vacuum at $\phi=-F/2m$ which using eq. (3) gives $\phi \sim -a \Lambda^3/2mM_P$. Note that for a positive $\mu$ we need $a<0$.
Since  $\lambda \phi= \mu$, a TeV scale $\mu$ requires (assuming $a \sim -1$ and the tree level mass $m \sim M_P$)
$\Lambda \sim 10^{-5}~M_P \sim 10^{13}~GeV$ which is a reasonable scale for a non--Abelian gauge group. 

This is a simple solution to the $\mu$--problem at the cost of introducing nonrenormalizable interactions with a hidden non--Abelian gauge group. In addition, above we assumed $a$ is small
and negative. The fact that, at tree level, the superpotential contains a mass term but no F--term is crucial for our result; 
if both terms are present at tree level (with a Planck scale), the VEV of $\phi$
is not exponentially suppressed but around $M_P$. The same is true if both terms vanish at tree level and arise due to nonperturbative effects as in eq. (2).
Unfortunately, in field theory, it is difficult to justify the presence of only the mass term. For example, R parity under which $R(H_u)=R(H_d)=1$ requires
$R(\phi)=0$ so both the $\phi$ mass and F terms break R parity. Thus R parity does not allow a tree level mass term and leads to a Planck scale VEV for $\phi$. It is easy to see that
discrete symmetries cannot lead to a nonzero mass term with a vanishing F--term either if we need the term in eq. (1).

We see that a small $\mu$--term can be obtained if we are willing to postulate the existence of a hidden non--Abelian gauge group with nonrenormalizable interactions to $\phi$. Even
more importantly, we need to assume that, at tree level, $W$ contains a mass term but no F--term for $\phi$. All this is quite unnatural in field theory and therefore the above mechanism is not a very
convincing solution.

\bigskip
\centerline{\bf 3. A Mechanism for Small $\mu$ in String Theory}
\medskip

In this section, we constuct a brane model which leads to a small $\mu$--term. This should not be considered to be a complete solution to the $\mu$--problem since the mechanism is
not embedded in a complete brane construction of the MSSM. Rather, we concentrate only on the physics that is relevant to the $\mu$--problem. This 
stringy mechanism is very similar to the one described in the previous section (in field theory) but it lacks the former's shortcomings. 

Consider the deformed $A_3$ singularity fibered over the complex plane $C(x)$ given by (this is essentially described in section 4 of ref. [\SHA])
$$uv=(z-mx)(z+mx)(z-mx)(z+m(x-2a)) \eqno(6)$$
The $A_3$ singularity has three nodes, i.e. $S_i^2$s, $i=1,2,3$. We wrap one D5 brane on $S_1^2$ and $S_3^2$ and two D5 branes on $S_2^2$. Thus the gauge group of the world--volume theory is 
$U(1)_1 \times U(2)_2 \times U(1)_3$.
The weak $SU(2)_W$ gauge group is the subgroup of $U(2)$ whereas the hypercharge $U(1)_Y$ is given by $[U(1)_1-U(1)_2]/2$. For every node ($S_i^2$) there is a gauge singlet $\phi_i$ and for every
link between the nodes there is a pair of fields in the bifundamental representation, $Q_{12},Q_{21},Q_{23},Q_{32}$[\QUI]. (We will later make the identification $Q_{12}=H_u$ and $Q_{21}=H_d$.)
The superpotential is given by the Yukawa terms which are inherited from the untwisted ${\cal N}=2$ supersymmetric $A_3$ singularity
$$W_1=(\phi_2-\phi_1)Q_{12}Q_{21}+ (\phi_3-\phi_2)Q_{23}Q_{32} \eqno(7)$$
Fibering $A_3$ over $C(x)$ leads to superpotential terms for the singlets given by[\SUP]
$$W(x_i)= \int(z_i(x)-z_{i+1}(x)) dx \eqno(8)$$
where $z_i(x)$ are the zeros of the different terms in eq.(6). This gives (identifying $x_i$ with $\phi_i$) the singlet superpotential
$$W_2=m\phi_1^2-m\phi_2^2+m(\phi_3-a)^2 \eqno(9)$$
The complete superpotential is $W=W_1+W_2$. $W$ has two parameters, $m$ and $a$ which are both assumed to be string scale. (In the following we will not distinguish between the string and Planck scales
since this is irrelevant for our purposes.) 
The superpotential does not contain any exponentially small parameters which can lead to a small VEV or
$\mu$. Moreover, since it is not corrected at the perturbative level, the only possible corrections are nonperturbative, e.g. due to instantons.

Since we relate the first two nodes ($S_1^2,S_2^2$) to the Standard Model gauge groups, the only instantons we can consider belong to the third node ($S_3^2$). The brane
instantons on this node are Euclidean D1 branes wrapping $S_3^2$. The effects of these instantons can be easily computed by a classical calculation after a geometric transition at this 
node[\GEO]. For this calculation to be reliable, the node has to be isolated and all the fields at the node must be massive. We see from eq. (6) that the first two nodes are at $x=0$ 
whereas the third one is at $x=a$. Thus the third node is isolated. Similarly from eqs. (7) and (9) we see that $\phi_3,Q_{23},Q_{32}$ are massive. As a result, we can trust the classical 
calculation of the superpotential after the geometric tansition.

Under the geometric transition, $S_3^2$ collapses to zero size and an $S^3$ blows up in its place. The D5 brane wrapping $S_3^2$ is replaced by one unit of Ramond-Ramond flux
$$\int_{S_3^2} H^{RR}=1 \eqno(10)$$
The new geometry is described by the equation
$$uv=(z-mx)(z+mx)((z-mx)(z-m(x-2a))-s) \eqno(11)$$
The size of $S^3$ is given by $S=s/m$. After the geometric transition, the fields that lived on the third node, $\phi_3,Q_{23},Q_{32}$, disappear. In addition, the transition gives rise 
to two corrections
to the superpotential. The first one is the nonperturbative superpotential for $S$[\NPS] (which describes the gaugino condensate for non--Abelian gauge groups) due to the flux in eq. (10) 
$$W_{flux}={t \over g_s}S+S \left( log{S \over \Delta^3}-1 \right) \eqno(12)$$
Here 
$$t=\int_{S^2}(B^{NS}+ig_s B^{RR}) \eqno(13)$$ 
$g_s$ is the string coupling and $\Delta$ is an ultraviolet cutoff around $M_s$. The second correction changes the superpotential for $\phi_2$ to[\SHA]
$$W^{\prime}(\phi_2)=\int^{\phi_2} (-m(y+a)-\sqrt{m^2(y-a)^2+s})dy \eqno(14)$$
Assuming $\phi_2<<a$ we get
$$W^{\prime}(\phi_2)=-m\phi_2^2-{1 \over 2} S~ Tr log \left( {{a-\phi_2} \over \Delta} \right) \eqno(15)$$
The superpotential for $\phi_1,Q_{12},Q_{21}$ remains as before. Now, we see that the fields $\phi_1,\phi_2, S$ are massive and can be integrated out leaving a superpotential for the 
light fields $Q_{12},Q_{21}$. Setting the F--terms for the heavy fields
$$F_{\phi_1}=2m\phi_1-Q_{12}Q_{21} \eqno(16)$$
$$F_{\phi_2}=-2m\phi_2+Q_{12}Q_{21}+{s \over {2(a-\phi_2)}} \eqno(17)$$
and $$F_s={t \over g_s}+log \left({S \over \Delta^3} \right)-{1 \over 2}Tr log{{(a-\phi_2)} \over \Delta} \eqno(18)$$
to zero, we can integrate them out. These F--terms vanish for (to lowest order)
$$S_0=\Delta^3 exp[(-(t/g_s)+(1/2)log(a/\Delta)] \eqno(19)$$ 
$$\phi_{10}=-{1 \over {2m}}Q_{12}Q_{21} \eqno(20)$$
and
$$\phi_{20}={1 \over {2m}}Q_{12}Q_{21}+{S_0 \over {4ma}} \eqno(21)$$
In order to calculate the instanton correction in eq. (15) we assumed $\phi_2<<a$ which is justified in light of eq. (21).
Substituting eqs. (19)-(21) into the superpotential for $\phi_1,\phi_2,S,Q_{12},Q_{21}$ we get the low energy superpotential for $Q_{12},Q_{21}$
$$W_{low E}={1 \over m}Tr(Q_{12}Q_{21}Q_{12}Q_{21})+{S_0 \over {4ma}}Tr(Q_{12}Q_{21}) \eqno(22)$$ 
The first term is negligible at low energies since $m \sim M_s$. The second term looks like a $\mu$--term if we can identify $Q_{12},Q_{21}$ with the Higgs fields $H_u,H_d$. $Q_{12}$ and $Q_{21}$ are
doublets under the $SU(2)$ subgroup of $U(2)$. Their charges under $U(1)_1 \times U(1)_2$ are $(1,-1)$ and $(-1,1)$ respectively. We see that the doublets are neutral under the 
combination $U(1)_1+U(1)_2$ and this Abelian group decouples. The charges of the doublets under $[U(1)_1-U(1)_2]/2$ on the other hand are $1$ and $-1$ and this Abelian group can be 
identified with hypercharge, $U(1)_Y$. Then, the electric charge is given by $Q=Y/2+T_3$.

From eqs. (19) and (22) for $S_0$ we see that the $\mu$ term is exponentially small compared to $M_s$,
$$\mu \sim {{\Delta^3 e^{-t/g_s}} \over {4ma}} \sim {1 \over 4}M_s e^{-t/g_s} \eqno(23)$$
where we assumed no fine tuning for the parameters, i.e. $\Delta \sim m \sim a \sim M_s$. A TeV scale $\mu$ requires $(t/g_s) \sim 35$ which can easily be accomodated by taking a large
enough $B^{NS}$ in eq. (13).

The brane model we described above is the string theory realization of the toy model of the previous section but without its shortcomings. For example, in the brane model the fields and
the superpotential (including the couplings) are not arbitrary but fixed by the underlying geometry. In addition, the instanton corrections to the superpotential are calculable due
to the duality related to the geometric transition. Thus, once a geometry is chosen with given a brane configuration, the brane world--volume theory is completely determined and
there is no freedom left as opposed to the situation in field theory which is quite arbitrary. 

We remind that, at tree level in field theory, we need a superpotential with a nonzero mass term and a vanishing F--term which cannot be justified through symmetries. From eq. (6)
that describes the geometry of the singularity, we see that the presence of an F--term for $\phi_1$ or $\phi_2$ is related to the location of the first two nodes in $C(x)$. In our case, 
the third node is at $x=a$ and the first two nodes are at $x=0$. Therefore, there is an F--term for $\phi_3$ ($\sim ma$) but not for $\phi_1$ and $\phi_2$. If the first two nodes are separate from 
each other (different VEVs for $\phi_1$ and $\phi_2$), we get tree level F--terms for $\phi_1$ and $\phi_2$, which leads to a string scale $\mu$. If the nodes coincide (away from the origin)  
we find that there is no tree level contribution to $\mu$ even in the presence of nonzero F--terms since the VEVs of $\phi_1$ and $\phi_2$ are equal. Of course, if the nodes coincide
at the origin as they do in our case there are no tree level F--terms or $\mu$. Thus, the geometry given by eq. (6) is more symmetric than the general case; it has two nodes that coincide 
at the origin of $C(x)$.
It is this extra symmetry of the geometry that is behind the world--volume theory without a tree level $\mu$.
The exponentially suppressed value of $\mu$ is a result of the nonperturbative corrections to the tree level superpotential which are given by the terms in eq. (12) and (15). The first
fixes the VEV of $S$ to be exponentially small. The second creates a coupling between $S$ and $\phi_2$ which leads to the $\mu$--term after the heavy fields are integrated out.

\bigskip
\centerline{\bf 4. Conclusions and Discussion}
\medskip

In this letter, we described a brane model on a deformed $A_3$ singularity which leads to an exponentially small $\mu$--term. The origin of the small $\mu$--term is the brane instanton correction to 
the superpotential which can be reliably calculated after a geometric transition at one of the nodes of the singularity. This requires that the singularity have a node that is isolated and the fields
living on it be massive. For the above mechanism to work, we also need two other coincident nodes of the singularity. Unless these nodes coincide in the complex plane $C(x)$ there
is a tree level F--term in the superpotential which leads to a string scale $\mu$. The model we described above, with three nodes, is the simplest one that satisfies all these requirements.
Clearly, there are many other possible models with branes wrapped on singularities which have the same properties and lead to a small $\mu$.

We saw that the origin of small $\mu$ is the specific geometry, i.e. the deformed and fibered $A_3$ singularity on which the D5 branes wrap. Since the $\mu$--problem is a problem of naturalness,
it is important to find out how natural the geometry that describes the singularity is. It may seem that we can always fix the coordinates of $C(x)$ so that the first node is at the origin.
This is not correct after we fiber the singularity, i.e. with the singlet mass terms in the superpotential since these fix the origin. We also need to locate the third node
away from the other two so that it is isolated and the instanton correction after the geometric transition can be calculated reliably. Therefore, the presence of 
a tree level F--term (for $\phi_1$ or $\phi_2$) depends on the location of the first two nodes. When the two nodes coincide (at or away from the origin) the symmetry of the singularity 
is enhanced. It is even more enhanced when the two nodes are at the origin which corresponds to the case we studied. This symmetry enhancement in the geometry seems to be responsible 
for the absence of an F--term in the superpotential which leads to a small $\mu$. Of course, the most symmetric case is the one
in which all three nodes coincide but, for our purposes, this is not desirable since we need to isolate one of the nodes from the others.

We did not embed the above stringy mechanism in a fully--fledged brane construction of the MSSM. This is much harder to do since a simple extension of our construction
to include quarks, leptons and the $SU(3)$ color gauge group does not seem to work. An alternative approach may be to try to realize the above mechanism in brane models that describe the MSSM on more
complicated singularities such as that in [\GW].

\bigskip
\centerline{\bf Acknowledgements}

I would like to thank the Stanford Institute for Theoretical Physics for hospitality.

\vfill

\refout

\end
\bye